# Cracking the code:
# Lessons from 15 years of digital health IPOs for the era of AI

**Tamen Jadad-Garcia**

Partner, Vivenxia Healthcare Consulting, Los Angeles, USA
tjadadgarcia@vivenxia.com

**Alejandro R. Jadad, MD DPhil LLC**

Founder, Centre for Digital Therapeutics, Toronto, Canada; Research Professor (Adjunct), Keck School of Medicine, University of Southern California, Los Angeles, USA
aj_492@usc.org or ajadad@gmail.com

## Abstract

**Introduction**: As digital health technologies continue to evolve, investors and industry leaders seek to identify the factors that drive success in this rapidly growing sector. This study addresses the question, *What separates the winners from the rest in this competitive landscape?* and examines the impact of billing codes for reimbursement on the long-term financial performance of digital health companies in U.S. stock markets.

**Methods:** Digital health companies that went public on U.S. stock exchanges between 2010 and 2021 were selected based on their offering of products or services aimed at improving personal health or disease management and their presence in the U.S. market. To identify eligible IPOs, a systematic search strategy using Google, Google News, and existing lists of IPOs was conducted. The companies were categorized based on the presence or absence of billing codes at the time of their IPO. We compared key performance indicators, including Compound Annual Growth Rate (CAGR), relative performance to benchmark indices, and market capitalization change, using Mann-Whitney U tests and Fisher Exact Tests.

**Results:** Of the 33 digital health companies analyzed, 15 (45.5%) had established billing codes at the time of their IPO, while 18 (54.5%) did not. The median IPO price for all companies was $17.00, with no significant difference between those with and without billing codes. Companies with billing codes were 25.5 times more likely to achieve a positive CAGR (p = 0.001). These companies had a median market capitalization increase of 56.3%, compared to a median decline of 80.1% for those without billing codes. Nine out of the 10 worst-performing companies, in terms of CAGR, lacked billing codes, while all five top performers had billing codes at IPO. Companies without billing codes were also 16 times more likely to experience a drop in market capitalization by the study's end (p = 0.001).

**Conclusion:** The impact of billing codes on the long-term success of digital health companies is far less intuitive than expected. Many founders, investors, and developers may have overestimated consumers' willingness to pay out-of-pocket or underestimated the complexity of securing institutional reimbursement. Market analysts and other key influencers in the investment landscape initially overlooked this crucial factor, potentially focusing more on hype, technological innovation, and short-term market dynamics. As the digital health sector evolves, especially with the rise of AI-driven solutions, stakeholders should prioritize billing codes as a strategic pathway to ensure sustainable growth, financial stability, and maximized investor returns.

# Introduction

As the U.S. healthcare landscape evolves with the advancement of digital technologies, so too does the race to innovate within it. Multinational corporations and investment groups have collectively poured billions into the development of digital health solutions, seeking to secure a share of a market projected to reach $6.8 trillion by 2030 (1).

With a rapidly aging population, a growing chronic disease burden, and persistent threats like pandemics and financial crises, the availability of affordable, scalable, and efficient healthcare solutions is more critical than ever. At the same time, the healthcare system faces a severe workforce crisis, with a shortage of more than four million workers by 2026 (2).

Digital health companies have emerged as critical players in addressing these challenges, offering innovative solutions that enhance accessibility, affordability, and quality of care (3). The shift from hospital-centric care to more distributed models, propelled by digital innovation, especially with the fast development of artificial intelligence (AI) creates unprecedented opportunities for health technology to play a vital role in supporting millions globally.

In 2020, the COVID-19 pandemic served as a catalyst, accelerating many pre-existing trends in healthcare technology. Investors flocked to the sector, with digital health funding surging to over $29 billion across 729 deals in 2021—nearly double the previous year's total (4). Private capital poured into companies at the intersection of healthcare and technology, with $13 billion invested in the first half of 2021 alone, compared to just $1 billion in all of 2012 (5).

Record investments, however, were followed by record bankruptcies. Healthcare bankruptcy filings in 2023 were more than three times higher than in 2021 (6). Venture funding for digital health in 2023 closed at the lowest levels since 2019, as investors pulled back amidst uncertainty (7).

In the midst of this turbulence, the emergence of generative AI has breathed new life into the sector. The potential for AI to revolutionize healthcare delivery has reignited investor confidence, signaling that the best times to transform the entire sector might be very close (8,9). However, given the rocky history of digital health companies in the marketplace, and the unprecedented level of uncertainty that accompanies generative AI, investors are very cautious, scrambling to judge whether and how much innovations are likely to be adopted, and how much business value they really offer (10).

As in any other sector, the key question in investors' minds is: *What separates the winners from the rest in this competitive landscape?*

The answer to this key question continues to be elusive, especially because of a surprisingly wide gap in the research literature on the factors that truly drive the success or failure of digital health companies, at all stages in their development, from startups to established ventures striving to scale up or to become financially sustainable (11).

This study seeks to address this gap by focusing on a previously overlooked but essential factor: the role that billing codes for the reimbursement of the procedures or services of digital health companies play on their financial performance. We have focused, specifically, on the impact of the presence or absence of such billing codes on the success or failure of digital companies at the time of their initial public offering (IPO). Our choice was driven by the hypothesis that having billing codes in place at the time of an IPO provides a competitive edge for a company in the market and significantly enhances its prospects for long-term success.

On the other hand, we considered IPOs as ideal for studying the impact of factors such as billing codes on the success or failure of digital health companies due to several unique characteristics. First of all, the IPO process requires companies to disclose a wealth of information, including financial statements, business models, market strategies, and regulatory compliance efforts, providing a rich dataset for analysis with a level of transparency that is often unavailable for private companies. In addition, IPOs are typically launched by companies that have reached a certain level of maturity, making them ideal cases for examining factors that are critical for long-term sustainability and financial performance. Finally, the transition to public markets acts as a real-world "stress test" that likely magnifies the effect of factors capable of influencing a company's ability to thrive in a competitive, high-stakes environment. Therefore, any valuable lessons learned from studying IPOs could be particularly valuable for investors, entrepreneurs, company executives, and developers involved in earlier stages of development or engaged in privately-owned initiatives, regardless of whether they intend to take them public.

## Methods

### Research question

What is the effect of the existence of billing codes aligned with the reimbursement of their products or services at the time of their IPO on the performance of digital health companies in the stock market?

### Study sample

Digital health companies that entered stock exchanges based in the U.S. from January 1, 2010 to December 31, 2021, with data on their performance for at least two years following their IPO.

### Selection criteria

Companies were regarded as eligible for inclusion in the study if they were listed as IPOs on NASDAQ or the NYSE; offered products or services that were driven by digital technology; aimed at improving personal health or disease management, including those for prevention, diagnosis, treatment, monitoring, and rehabilitation (12); targeted customers in the U.S.; and went public between January 2010 and December 2021.

IPOs were excluded from the study if they were Special Purpose Acquisition Companies (SPACs); had merged or been acquired; had undergone reverse listing; had engaged in follow-on a public offer (FPO) or secondary offering; had experienced extreme, unexplained, extraordinarily anomalous stock price movements; or if their ticker had been retired from the corresponding exchange. Companies were also excluded if they offered digital tools for activities unrelated to the direct provision of healthcare services to healthy people or patients; if digital technologies were included in their products or services as a secondary component of a primary medical device, rather than to influence personal health or disease management directly; or if they only offered services that were indirectly related to personal health or disease management (e.g., clinical research tools, insurance offerings, or administrative or financial support).

### Search strategy

We searched the web using Google and Google News with the following Boolean string:

> *[IPO OR "initial public offering"] AND [intitle:"digital health" OR intitle:"medtech" OR intitle:"med tech" OR intitle:"digital medicine" OR intitle:"medical device" OR intitle:"medical technology" OR intitle:"digital therapeutics" OR intitle:"digital diagnostic" OR intitle:"digital diagnosis" OR intitle:"health technology" OR intitle:"healthtech" OR intitle:"health tech" OR intitle:"digital medicine" OR intitle:"healthcare technology" OR intitle:telehealth OR intitle:telemedicine]*

We screened, independently, the first 200 links generated by the search engines, and used the information on the title and home page of any site that included details of any potentially eligible IPO.

We also screened existing lists included in key articles, reports and news feeds focused on IPOs in digital health (7,13–15).

After completing the screening process, we met and compared our assessments. Whenever there was disagreement about the eligibility of any particular IPO, the discrepancy was resolved by consensus.

**Data Collection**

Each of the selected IPOs was reviewed independently by each of us, independently, to verify their compliance with the inclusion criteria. If there were disagreements, we resolved them through dialogue, by consensus.

Once the sample was assembled, we extracted the following data from each of the selected IPOs:

- Company name
- Company ticker
- Headquarter location (US or elsewhere)
- Stock exchange (NASDAQ or NYSE)
- Background of the CEO (healthcare or not)
- Digital health domain (disease prevention, disease diagnosis, disease treatment, organization of care, rehabilitation)
- Existence of at least one billing code for reimbursement of the company's procedures or services on IPO date
- First day of trading (IPO date) at any point between January 1, 2010 and December 31, 2021
- IPO price
- Opening price
- First day closing price
- Lowest historical price
- Benchmark index value at closing of first day, and at every anniversary, and on June 30, 2024, which marked the end of the analysis period
- Stock price at each anniversary of the IPO up to June 30, 2024
- Market capitalization at the closing of first day and on June 30, 2024
- Composite index value at each of the stock's anniversaries and on June 30, 2024
- Number of months from IPO to June 30, 2024

The data about each of the companies and their stock values, and the composite indices were obtained from publicly available databases, namely Yahoo Finance and Stock Analysis.

The existence of relevant billing codes was determined by screening the website of the companies, by reviewing publicly available materials they had submitted to the Securities and Exchange Commission before going to market and by scanning news reports within a week of the IPO date. A company was deemed to have a reimbursement code for its procedures or services if, in any of the above sources, a Current Procedural Terminology (CPT) code was mentioned, or if there was an explicit statement about the eligibility of the company's procedures or services for reimbursement by at least one insurance company in the U.S., or by the U.S. Centers by Medicare and Medicaid Services (16).

The information gathered was stored in tabular format using Microsoft Excel (version 16.75.2).

**Data analysis**

Using the collected data, we calculated the following performance indicators for each of the IPOs (expressed as percentages, until noted otherwise):

*Intraday Price Change* = [(First Day Closing Price - Opening Price) / Opening Price] * 100

*Opening Premium/Discount* = [(Opening Price - IPO Price) / IPO Price] * 100

*Closing Premium/Discount* = [(First Day Closing Price - IPO Price) / IPO Price] * 100

*Compound Annual Growth Rate (CAGR)* = [(Stock Price at End / Stock Price at Start) ^ (1 / n)] - 1

*Price Return* = [(Stock Price at End - Stock Price at IPO) / Stock Price at IPO] * 100

*Relative Performance to Benchmark* = Price Return - Cumulative Benchmark Return

*Market Capitalization Change* = [(Market Cap at End - Market Cap at IPO) / Market Cap at IPO] * 100

At the outset, before completing any of the analyses, we considered the CAGR as the primary KPI, with the Relative Performance to Benchmark and the Market Capitalization Change as secondary KPIs.

The continuous values of each of the KPIs for the companies with a billing code for reimbursement of their products or services by the time of their IPO were compared with the counterparts that lacked such a code. The comparisons were conducted using Mann-Whitney U tests, with a p-value < 0.05 as the threshold for statistical significance.

In addition, we also split the values of Intraday Price Change, Closing Premium/Discount, Price Return, CAGR and Relative Performance to Benchmark into two groups, depending on whether they were positive or negative. We also noted companies whose stock price dropped more than 90% at any of the post-IPO anniversaries or on June 30, 2024, and whether their stock price on June 30, 2024 was greater than price at closing on IPO date. We used such data to compare companies with a billing code for reimbursement of their products or services by the time of their IPO with their counterparts, using a Fisher Exact Test. Odds ratios and 95% confidence intervals were calculated when the differences between the groups were statistically significant, using as a threshold also a p-value below 0.05.

Finally, we ranked the companies in descending order according to their CAGR, and selected the best and worst five performers. Relative Performance to Benchmark and Market Capitalization Change were used to break ties or further refine rankings when companies had similar CAGRs, and also to provide additional context and differentiate between companies when necessary.

# Results

**General characteristics**

The search led to the identification of 33 that met all of the inclusion criteria. Sixty companies were excluded (Table 1).

Of the 33 included companies, 30 (90.9%) were based in the United States at the time of their IPO, 25 (75.8%) were listed on the NASDAQ, and 25 (75.8%) had CEOs with a background in healthcare at the time of their IPO (Table 2). In terms of digital health domains, 10 companies (30.3%) focused on the organization of healthcare delivery, 9 (27.3%) on disease treatment and 5 (15.1%) on diagnosis, while the rest offered monitoring, telehealth, prevention or rehabilitation services. Eight (24.2%) of the IPOs took place between 2010 and 2015, 17 (68.5%) between 2016 and 2020, and 8 (24.2%) of them in 2021, with none in 2011 or 2012.

The included companies entered the public market arena with IPO prices averaging $15.90, ranging from $3.10 to $33.00. A median price of $17.00 suggests a balanced approach to valuation, with half of the companies pricing between $10.50 (25th percentile) and $20.50 (75th percentile).

Of the 33 companies, 15 (45.5%) had an established billing code on their IPO date. There was no statistically significant difference in the IPO price between those companies that had and did not have billing codes (p = 0.82) (Table 3).

**First-Day Performance**

On the first day of trading, 28 companies (84.8%) experienced a positive initial return, indicated by a closing premium, as their closing prices were higher than their IPO prices. Three companies (9.1%) had a negative initial return, reflected by a closing discount, with closing prices below their IPO prices. Two companies (6.0%) had a zero initial return, with closing prices equal to their IPO prices. These figures are not statistically significantly different (p = 0.2) from those of a sample of 2,111 IPOs in the United States from 2000 to 2020, in which 73.1%, 21.3% and 5.6% of the companies had positive, negative and zero first-day returns, respectively (17).

The median first-day return of the 33 IPOs was 39.8%, which is four times larger than what has been reported elsewhere for a general sample of 2,111 IPOs, estimated at 10.0% (17).

The median market capitalization reached by the companies at the end of the first day of trading was $526 million, with values ranging from a minimum of $15.0 million to a maximum of $12.7 billion. Overall, 15 (45.5%) of the companies exceeded $1 billion in market capitalization by the end of the first day of trading. This included only one of the companies with extraordinarily anomalous closing premiums.

The closing prices, the first day stock return, the intraday price change or the market capitalization of the companies with a billing code by their IPO date were not statistically significantly different than those without a billing code (p = 0.13, 0.60, 0.91 and 0.90 respectively).

**Long-term performance**

*Stock price patterns*

Only three of the 33 companies (9.1%)—Inspire Medical Systems, iRhythm Technologies and American Well—had stock prices that remained above the closing value at IPO date at each of the subsequent anniversaries. Of the remaining 30 companies, 17 (51.5%) had stock prices that stayed below the closing price at IPO date throughout the entire period of the study, with the remaining 13 (39.4%) showing an inconsistent pattern.

By June 30, 2024, 10 (30.3%) of the 33 companies had stock prices that were above their closing price on their IPO date. Nine of these companies had a billing code for their reimbursement of their services or products by their IPO date. This implies that digital health companies with billing codes were 25.5 times more likely to have a stock price above their IPO price by June 30, 2024, compared to companies without billing codes (95% CI of the odds ratios from 2.65 to 245.84, p = 0.0015).

Another 11 (33.3%) companies saw a drop in their stock price greater than 90% by June 30, 2024. Nine of them lacked a billing code for reimbursement at the time of their IPO, making the difference with those without a billing code statistically significant (p = 0.03).

*Compound Annual Growth Rate (CAGR)*

Ten companies achieved a positive level of CAGR by the end of the study period. When ranked from highest to lowest levels of CAGR, the top nine had billing codes. Only one of the 18 companies without a billing code, Evolent Health, had positive CAGR. This made companies with billing codes 25.5 times more likely to experience positive growth than those that lacked a billing code at the time of their IPO (p = 0.001).

Of the bottom 10 IPOs in terms of their CAGR, nine lacked billing codes at the time of their IPO. The only one that did was the worst performer of all 33 companies, American Well, which offers telehealth-based urgent care services.

*Relative performance to benchmark*

Five (15.1%) of the companies outperformed their corresponding benchmarks. This proportion is not significantly different from that of companies in the S&P 500, where only 22% IPOs outperformed the index from 2000 to 2010 (p = 0.39) (18).

These five companies had billing codes at the time of their IPO, while none of the 18 without billing code did (p = 0.012). All five also had positive CAGR levels.

*Price Return*

The same 10 companies that had positive CAGR had positive price returns by June 30, 2024. Companies with a billing code had significantly higher price returns as compared with their counterparts without a code, with medians of 49.0% and -75.4%, respectively (p = 0.0008).

*Market capitalization*

By June 30, 2024, 13 companies (39.4%) had market capitalization above $1 billion, of which 9 (69.2%) had achieved unicorn status at closing on IPO date. The length of time that the latter maintained this status ranged from 3 years to 9 years.

Digital health companies with billing codes at IPO achieved a median market capitalization change of 56.3%, with top performers exceeding 1000%. Those without billing codes at IPO experienced steep declines, with a median market capitalization ratio of -80.1%, with one third of them losing more than 90% of their value by the end of the follow up period.

Of the 15 companies with billing codes at IPO date, five (31.2%) experienced a drop in their market capitalization by June 30, 2024, while 16 of the 18 companies without a billing code at the outset saw their market capitalization shrink by this point. The difference was statistically different (p = 0.001)

As a whole, companies without a billing code at the time of their IPO were 16 times more likely to experience a drop in market capitalization compared to those with a billing code (p value = 0.001).

**Overall performance**

*Best performers*

The five companies with positive CAGR and stock returns, and that exceeded the performance of their corresponding indices TransMedics Group, Inspire Medical Systems, iRhythm Technologies, Semler Scientific and Electromed (Table 4).

All of these companies had billing codes by the time of their IPO, and had market capitalizations that grew from more than 200% since their IPO date (Semler Scientific) to just over 1000% (TransMedics Group).

In general, top performers seemed to have a more focused approach than the rest, specializing in medical devices or diagnostic tools targeting specific health conditions. This specialization might have contributed to more predictable revenue streams, established reimbursement models, and less competitive pressure, leading to their overall financial stability and growth.

*Worst performers*

Five of the companies with the lowest levels of CAGR had stock prices that had dropped more than 90% by June 30, 2024. These companies, which also had relative performance to benchmark levels lower than -129.2%, included, in ascending order, American Well, Renalytix, Moyano, NantHealth and Outset Medical (Table 4).

American Well was the only company in this group with billing codes that enabled the reimbursement of its services. Similarly to the other two companies offering telehealth services in the sample—Doximity and Teladoc—it did not outperform the market index (the NYSE Composite) during any of the anniversaries of their IPO, and its stock price on June 30, 2024 was below the IPO price. All three companies, on the other hand, saw a peak in their performance during 2021, reflecting the favorable conditions that the COVID-19 pandemic created for remote healthcare services.

As a whole, the worst performing companies appeared to have broader, less focused strategies; operated in highly competitive and crowded markets; and might have been more vulnerable to shifts in investor sentiment than their top performing counterparts.

## Discussion

Our analysis uncovered a key, yet previously overlooked, factor that has an important impact on the long-term success of digital health companies: the presence of billing codes for reimbursement. Companies with billing codes were more likely to achieve positive CAGR and outperform their benchmark indices over time, and were 25.5 times more likely to have a stock price above their IPO price by June 30, 2024, compared to those lacking billing codes. Remarkably, all five of the top performers in terms of CAGR and relative performance to benchmark had established billing codes by the time of their IPO, while only one of the companies without billing codes achieved positive CAGR. Moreover, companies without billing codes at the time of their IPO were 16 times more likely to experience a decline in market capitalization by the end of the study period.

While these findings may seem intuitive, our study revealed that the effect of billing codes on digital health company performance has been far less obvious than expected. Notably, more than half of the companies that entered the market from 2010 to 2021 did so without an established mechanism for their procedures or services to be reimbursed by insurers or government programs. This fact suggests that many founders, company executives, investors, and developers may have overestimated the readiness of individual consumers to pay out-of-pocket for services, or underestimated the complexity and time involved in opening new pathways to secure institutional reimbursement.

The strong influence of billing codes on the long-term performance of digital health IPOs could be attributed to their role as a critical pathway for securing consistent and scalable revenue streams. Billing codes serve as the bridge between healthcare providers and payers, including insurance companies and government programs, enabling reimbursement for the use of digital health services or products. Without these codes, companies face significant challenges in generating predictable income; with them, they have a competitive edge in the market and greater prospects for long-term success.

### Practical implications

Our findings have several practical implications. First, they underscore the importance for company founders, executives and developers to optimize their offerings for reimbursement, ensuring they meet the criteria for billing codes from the outset. The powerful effect of billing codes on long-term financial success for digital health companies should motivate investors to consider them as a key criterion when evaluating potential investment opportunities. In other words, billing codes, by default, validate the company's business model, while reducing the risk of failure by demonstrating that there is a clear pathway to generate revenue within the healthcare system. In addition, the evidence from our study serves as a crucial guide for stakeholders across all stages of digital health company development, emphasizing that reimbursement pathways are not just a concern for established firms but also for those at much earlier stages, from startups, through private companies in growth phase, all the way to exit or acquisition.

Another key observation from our study with important practical ramifications is that there were no significant differences in IPO performance during the first day of trading between companies with and without billing codes. This finding suggests that analysts, traders, and other stakeholders who influence market dynamics may have initially overlooked the importance of billing codes as a critical determinant of long-term financial success. It implies that the presence of billing codes was not a primary factor in their immediate valuation or investment decisions, despite its clear impact on the companies' sustainability and growth prospects over time. This oversight could be attributed to a lack of awareness or understanding of how integral reimbursement pathways are to the revenue generation and profitability of digital health companies. As a result, early market perceptions may have been driven more by hype, technological innovation, or broader market trends, rather than by a thorough assessment of the companies' potential for securing sustainable revenue streams through established reimbursement mechanisms.

The practical implications of our study could be even more significant for those seeking to capitalize on the rapidly evolving landscape of AI in healthcare, following the introduction of the "Taxonomy of Artificial Intelligence for Medical Services and Procedures" into the CPT code set in 2022 (19). This taxonomy has been proposed by the American Medical Association (AMA) to act as a structured framework to help categorize AI services into three distinct types: "Assistive," "Augmentative," and "Autonomous," based on their interactions with healthcare professionals.

"Assistive" AI detects clinically relevant data but still requires physician interpretation. It may be viewed as an extension of existing workflows, offering enhancements to efficiency and decision-making without replacing human judgment.

"Augmentative" AI solutions, in contrast, are those that provide more advanced analytics or quantification of data. They offer clinically meaningful insights that significantly influence patient care, but still require a physician's interpretation.

The "Autonomous" category encompasses solutions capable of independently generating clinical conclusions and recommendations. These represent the most significant potential shift in how healthcare services are delivered.

Each category carries distinct implications for how a product, service or procedure would be valued, adopted, and reimbursed, based on its clinical contribution, complexity, and financial impact.

- "Assistive" AI may face fewer adoption barriers and be easier to reimburse, as they can be positioned as enhancements to established procedures with existing billing codes. On the other hand, they may also command lower reimbursement rates as they would not significantly reduce physician workload.

- "Augmentative" AI solutions will likely be valued higher due to their greater contribution to clinical outcomes. However, securing reimbursement for them might be more complex, as companies will need to demonstrate that the AI-generated insights translate into measurable and safe improvements in patient care and efficiency.

- "Autonomous" products have the potential to revolutionize care by reducing the need for direct human intervention, but they will also face the greatest scrutiny in terms of regulatory approval, adoption, and reimbursement. However, they also have the potential for the highest valuation and reimbursement rates, particularly if it can be demonstrated that they can substantially improve patient outcomes or reduce costs.

This differentiation may also steer innovation in specific directions, as companies aim to develop AI solutions that align with the categories that offer the most favorable combination of adoption potential and reimbursement rates.

## Limitations

While our study provides valuable insights into the critical role of billing codes in the long-term success of digital health companies and the emerging opportunities for AI-driven solutions, several limitations should be acknowledged.

Firstly, although our sample size of 33 companies might seem relatively small, it already represents the largest collection of digital health IPOs during the specified timeframe. Nevertheless, this limited sample size may affect the generalizability of our findings to the broader digital health landscape, particularly given the rapid pace of technological advancement and regulatory changes in the healthcare sector.

Secondly, our study focused exclusively on companies that went public between 2010 and 2021. While this period captures an important phase in the evolution of digital health, it does not fully encompass the most recent developments, particularly the surge in AI-driven healthcare solutions that have emerged since 2022. The introduction of the AI taxonomy in the CPT code set, which occurred after our study period, may have already begun to influence how companies approach reimbursement strategies, suggesting that a subsequent analysis could yield even greater insights.

Thirdly, our analysis was limited to publicly traded companies. While this approach provided access to standardized financial data, it may not have captured the full spectrum of digital health innovation, much of which occurs in privately held companies or early-stage startups. The dynamics of securing billing codes and their impact on success may differ for these non-public entities, which could present different challenges or opportunities in achieving financial sustainability.

Fourthly, while we identified a strong correlation between the presence of billing codes at IPO and long-term financial success, our study design does not allow us to establish causality. Other factors, such as product quality, market conditions, regulatory environment, or management expertise, may also play significant roles in a company's performance.

Lastly, our study did not differentiate between different types of billing codes or reimbursement pathways. Given the complexity and variability of reimbursement mechanisms across different healthcare settings and payers, this could have influenced the impact of billing codes in ways that our study did not capture.

These limitations, while important to acknowledge, also highlight exciting avenues for future research, underscoring the need for ongoing studies that can adapt to the dynamic nature of digital health, particularly as AI technologies continue to reshape the sector.

## Concluding remarks

Our study has uncovered a pivotal factor in the success of digital health companies: the presence of billing codes for reimbursement. This finding underscores that sustainable growth in this sector requires more than technological innovation; it demands strategic alignment with the financial and regulatory frameworks that underpin the healthcare system. As we move further into an era where AI is increasingly central to healthcare delivery, the importance of integrating reimbursement pathways from the outset cannot be understated.

The recent introduction of the AI taxonomy into the CPT code set represents a landmark development, offering a structured pathway for AI-driven solutions to become an integral part of the healthcare ecosystem. This presents a unique opportunity for digital health companies, investors, and innovators to navigate this evolving landscape with greater clarity and purpose. By understanding how AI solutions can fit into established billing frameworks, stakeholders can more effectively support the adoption and scaling of technologies that have the potential to transform healthcare delivery.

For investors and stakeholders considering large-scale initiatives, our findings offer guidance that can inform future investments and strategic decisions. Companies that prioritize reimbursement pathways are not only more likely to achieve financial sustainability but are also better positioned to contribute to a more efficient, accessible, and equitable healthcare system. This creates a compelling case for a deeper engagement with digital health ventures that have demonstrated a commitment to aligning their offerings with the reimbursement mechanisms that are essential for long-term success.

While our study has limitations, it opens up exciting avenues for future research, particularly in understanding how reimbursement models will continue to evolve alongside advancements in AI and digital health. As we continue to explore this dynamic field, we look forward to collaborating with others who share our vision for a healthcare future that harnesses the full potential of digital innovation, ensuring that the benefits of these technologies reach as many people as possible.

In this journey, we invite fellow researchers, industry leaders, and investors to join us in advancing knowledge and driving meaningful change in the digital health sector, working together to build a foundation for innovations that are not only groundbreaking but also financially and operationally sustainable in the long term.

# Tables

### Table 1. Excluded companies

| Reason for exclusion* | Companies |
|---|---|
| SPACs (n = 19) | 23andMe, Augmedix, Babylon, Beachbody Company, Butterfly Network, Cano Health, Clover Health, DocGo, GeneDx, Hims & Hers Health, Hyperfine, LumiraDx, Owlet Baby Care, P3 Health Partners, Pear Therapeutics, Sharecare, Talkspace, UpHealth, Vicarious Surgical |
| Indirect services to patients or clinicians (n = 17) | Alignment Healthcare, Computer Programs & Systems, Definitive Healthcare, GoHealth, Health Catalyst, HealthStream, Iqvia, Marpai, Omnicell, Oscar Health, Phreesia, Progyny, SCWorx Corp, Streamline Health Solutions, Veeva Systems, Veradigm, Weave |
| Merged or acquired, with ticker retired from exchange (n = 9) | Convey Health Solutions, Fitbit, Livongo, Oak Street Health, Ortho Clinical Diagnostics, Science 37, Signify Health, SOC Telemed, Tabula Rasa HealthCare Inc. |
| Medical devices with digital tools as support (n = 8) | DexCom, IRadimed, Myomo, Nevro Corp, NovoCure, Pulse Biosciences, Talis Biomedical, Vivos |
| Biotechnology/drug discovery companies (n = 3) | Schrodinger, Simulations Plus, Sophia Genetics SA |
| Delisted from exchange (n = 2) | Better Therapeutics, Cue Health |
| Follow-on public offer or secondary offering (n = 1) | R1 RCM |
| Extreme, unexplained, extraordinarily anomalous price movements (n = 1) | Avinger (Jumped from an IPO price of $18 to a closing price of $1,578,000) |

*As more than one category might apply, the most dominant was selected

**Table 2. General characteristics of the included companies**

| | |
|---|---|
| **Total number** | 33 |
| **Geographic base** | USA: 30 (90.9%)<br>Israel: 3 (9.1%) |
| **Digital Health Domains** | Organization: 10 (30.3%)<br>Treatment: 10 (30.3%)<br>Diagnosis: 5 (15.2%)<br>Telehealth: 3 (9.1%)<br>Monitoring: 3 (9.1%)<br>Prevention: 1 (3.0%)<br>Rehabilitation: 1 (3.0%) |
| **Exchanges** | NASDAQ: 25 (75.8%)<br>NYSE: 8 (24.2%) |
| **Year of IPO** | 2021: 8 (24.2%)<br>2020: 5 (15.1%)<br>2019: 3 (9.1%)<br>2018: 5 (15.1%)<br>2017: 1 (3.0%)<br>2016: 3 (9.1%)<br>2015: 2 (6.1%)<br>2014: 4 (12.1%)<br>2013: 1 (3.0%)<br>2010: 1 (3.0%) |

**Table 3. IPO Price**

| | Companies with billing code (n = 15) | Companies without billing code (n = 18) |
|---|---|---|
| **Mean** | 15.30 | 16.40 |
| **Standard Deviation** | 5.4 | 9.2 |
| **Median** | 17.00 | 16.00 |
| **Minimum** | 4.00 | 3.10 |
| **25% Percentile** | 13.00 | 7.60 |
| **75% Percentile** | 18.00 | 23.00 |
| **Maximum** | 26.00 | 33.00 |
| **Range** | 22.00 | 29.90 |

**Table 4. Best and worst performing IPOs, ranked**

| Ranking | Company name | Exchange | IPO date | CAGR (%) | Relative Performance to Benchmark (%) | Price Return (%) | Market Capitalization Change (%) |
|---|---|---|---|---|---|---|---|
| 1 | **TransMedics Group** | NASDAQ | 5-2-19 | 56.2 | 708.6 | 831.1 | 1050.5 |
| 2 | **Inspire Medical Systems** | NYSE | 5-3-18 | 42.5 | 694.8 | 739.1 | 688.3 |
| 3 | **iRhythm Technologies** | NASDAQ | 10-20-16 | 24.6 | 239.0 | 480.1 | 483.1 |
| 4 | **Semler Scientific** | NASDAQ | 2-21-14 | 16.7 | 48.5 | 367.9 | 643.8 |
| 5 | **Electromed** | NYSE* | 8-13-10 | 9.4 | 88.3 | 250.0 | 266.7 |
| 29 | **Outset Medical** | NASDAQ | 9-15-20 | -38.9 | -145.8 | -86.0 | -82.1 |
| 30 | **NantHealth** | NASDAQ | 6-2-16 | -53.6 | -359.4 | -99.8 | -99.2 |
| 31 | **Movano** | NASDAQ | 3-23-21 | -60.9 | -129.2 | -94.0 | -80.4 |
| 32 | **Renalytix** | NASDAQ | 7-17-20 | -63.0 | -168.4 | -98.1 | -96.0 |
| 33 | **American Well** | NYSE | 9-17-20 | -64.2 | -142.0 | -98.4 | -98.1 |

* NYSE American